\documentclass{webofc}
\usepackage[varg]{txfonts}   
\usepackage{graphicx,wrapfig,lipsum}
\begin{document}
\title{Proton-cluster femtoscopy with the HADES experiment}
%
%

\author{\firstname{Maria Stefaniak} \lastname{for HADES Collaboration}\inst{1,2}\fnsep\thanks{\email{stefaniak.9@osu.edu}}
}

\institute{Department of Physics, The Ohio State University, Columbus, Ohio 43210, USA; 
\and
GSI Helmholtz Centre for Heavy Ion Research, 64291 Darmstadt, Germany
}

\abstract{%
 The matter created in Ag+Ag collisions at $\sqrt{s_{NN}}$ = 2.55 GeV, as measured with the HADES experiment, can be characterized by similar thermodynamic quantities as Neutron Star Mergers, thus becoming an essential reference for the understanding of these compact stellar objects. One of the methods applied to investigate heavy-ion collisions  are femtoscopic correlations. They are a unique tool for the determination of the interactions between hadrons and allow to search for possible exited or unbound states of nuclear matter. We performed precise experimental studies of the correlations between protons and different clusters and compared them with the existing theoretical descriptions.
}
\maketitle
\section{Introduction}
\label{intro}
The exploration of the QCD phase diagram is the goal of many nuclear facilities. Recently, the region of finite net-baryon densities and relatively low temperatures received much experimental and theoretical interest. It is expected that the study of heavy-ion collisions in this region can serve as a reference to astrophysical studies of Neutron Star Mergers (NSM), as they are crucial for the determination of the Equation of State (EoS) of nuclear matter. Strong Interactions (SI) between hadrons and exited or unbound states of nuclear matter are one of the still not well described elements required to understanding the EoS. They can be studied with the HADES experiment using a tool called Femtoscopic Correlations (FC). 

\section{HADES experiment} \label{hades}
\textbf{H}igh \textbf{A}cceptance \textbf{D}i\textbf{E}lectron \textbf{S}pectrometer is a fixed-target experiment at the SIS-18 accelerator located at the GSI Helmholtz Centre for Heavy Ion Research in Germany. It has an almost full azimuthal acceptance and covers polar angles between $18^0$ and $85^0$. For this study, the data collected on collisions of Ag+Ag at $\sqrt{s_{NN}}$ = 2.55 GeV are studied. The created systems can be described by temperatures ($T$ = 60-80 MeV) and baryon densities ( $\rho$ < 2-3 $\rho_0$) which is close to ones characterizing NSM \cite{HADES:2019auv}.

\section{Femtoscopic correlations}
\label{sec:femto_corr}

FC probe the geometric and dynamic properties of the particle emitting source \cite{FemtoLisa}. They provide information about possible excited states of particles and allow to extract information on the SI between the involved particles. They can be described with the Koonin-Pratt formula: 
\begin{equation}
    C(k^{\star}) = \int S(r^{\star})| \Psi(k^{\star},r^{\star})|^2 d^3r^{\star} 
\end{equation}
where:
\\ $S(r^{\star})$ is the \textit{source function}, describing the distribution of the relative positions of the two particles ($r^{\star}$),
\\ $\Psi(k^{\star},r^{\star})$ is the \textit{two-particle wave function}. It is derived from the Schrödinger equation for a given potential characterizing the interactions between the particles,
\\ and $k^{\star}$ is the momentum of the particles in the pair rest frame. 
The shape of the CF is driven by the convolution of three effects:
\begin{itemize}
    \item Coulomb interaction, depending on the electric charge of particles,
    \item Quantum statistics, which is repulsive for fermions due to Pauli exclusion law,
    \item SI depending on the potentials between correlated particles.
\end{itemize}

Experimentally, the femtoscopic correlation function is based on the relative momenta of pairs of particles and can be expressed as: $C(k^{\star}) = A(k^{\star}) / B(k^{\star})$, where $A(k^{\star})$ and $B(k^{\star})$ are the distributions of $k^{\star}$ of all possible pairs within one event and mixed events, respectively. The latter one represents the uncorrelated particles, so introducing the division helps in suppression of non-femtoscopic correlations originating from for example flow, non-uniform detector efficiency, etc. 

\section{Results and discussion}

We performed a systematic study of the proton-cluster CF starting from the \textit{p-p} system, providing the baseline for more complex subsequent ones. 
The measurements were extended by adding an extra neutron to the composition of the cluster, which allows the visualization of the impact of additional neutrons on the obtained correlation functions. As neutrons do not carry any electric charge, the Coulomb force does not vary from system to system. The significant difference between the correlation functions is expected to originate therefore only from the SI.

\begin{wrapfigure}{l}{7cm}
\includegraphics[width=0.5\textwidth]{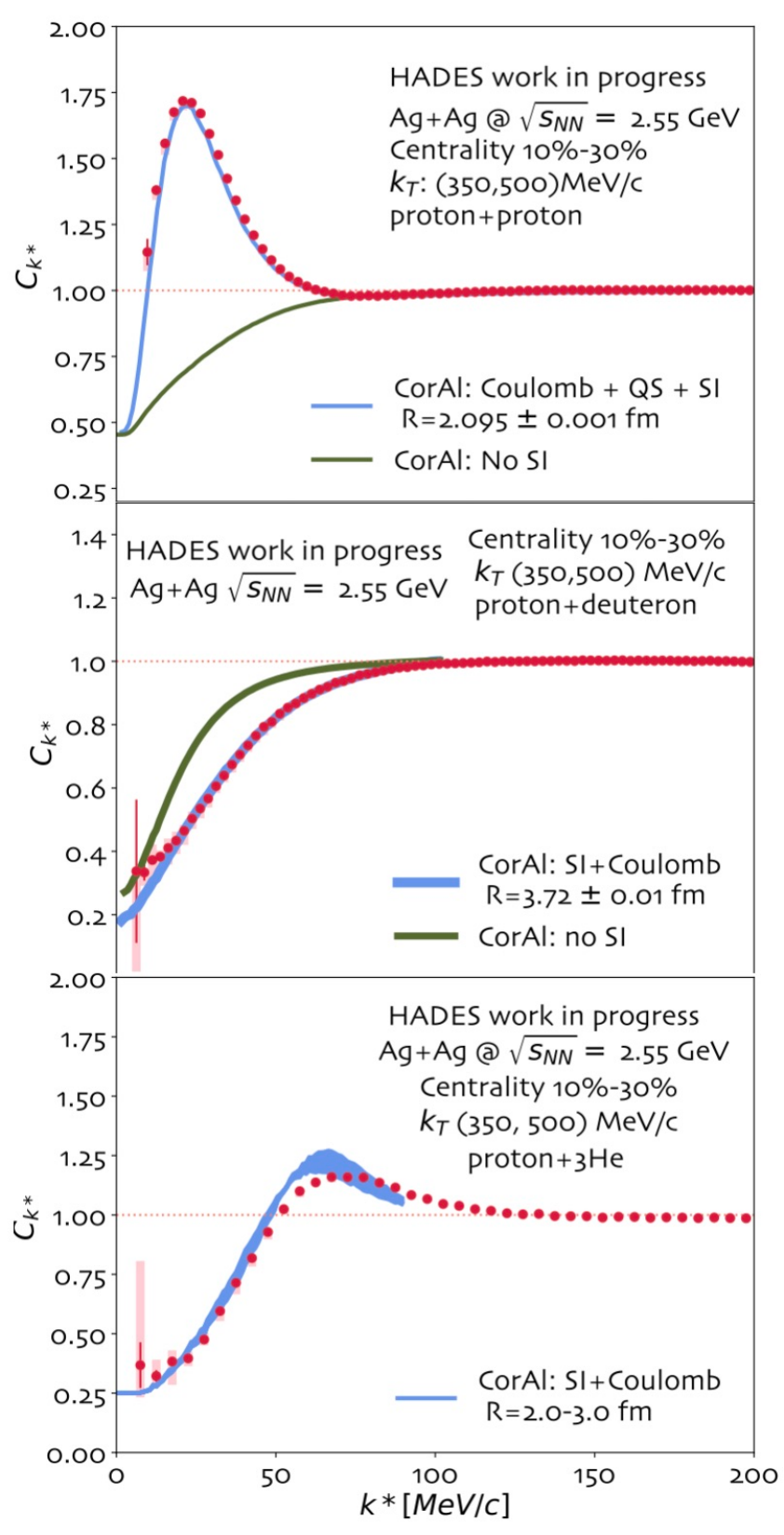}
\caption{Proton-cluster correlation functions for Ag+Ag collision at $\sqrt{s_{NN}}$ = 2.55 GeV for the centrality  $10\%-30\%$. The blue curves correspond to the full theoretical description and the green one does not consider SI.}
\label{fig::cf_pcluster}
\end{wrapfigure}

Figure \ref{fig::cf_pcluster} shows the set of CF starting from the top panel: proton-proton, proton-deuteron, and proton - $^3$He. In order to take care of two-track effects (merging) several cuts were applied: the difference between the $\theta$ (azimutal) angle between particles >0.04 rad, while the $\Delta \phi$ > 0.25 rad. Still, track merging is the main source of the systematic uncertainties. The correction for energy loss and momentum resolution are not yet applied in the presented data, but will be included in near future.
The experimental data are compared with theoretical predictions calculated with the CorAl software \cite{coral}. For all the systems the SI are known from previous measurements and studies:
\begin{itemize}
\item for p-p correlation the soft core Reid potential is used \cite{pp}, 
\item in p-d potential calculations the phase shifts from \cite{pd} are applied. However, only the s-wave is considered here. The calculations are going to be extended with the p-wave in the near future.
\item for p-$^3$He the phase shifts from \cite{p3he} are used, here the s- and p- waves were considered.
\end{itemize}

The theoretical predictions for both p-p and p-d, are in good agreement with the experimental data. The green curves originate from calculations without including the SI. In the case of p-p CF, the substantial attractive correlation is clearly driven by SI. On the other hand, the SI in p-d increases the repulsion between these two particles. For p-$^3$He the theoretical predictions for radii in the interval 2.0-3.0 fm are shown, as the fit cannot be performed. The maximum of CorAl CF is observed at lower $k^{\star}$ in comparison to the HADES data. Adding d-wave contributions to the calculations could potentially improve the theoretical description.
The changes in the shapes of the CF between the systems differentiated by additional nucleons are significant. The SI introduces essential variations from the Coulomb-driven interactions. 

\begin{figure}
\centering
\includegraphics[scale=0.5]{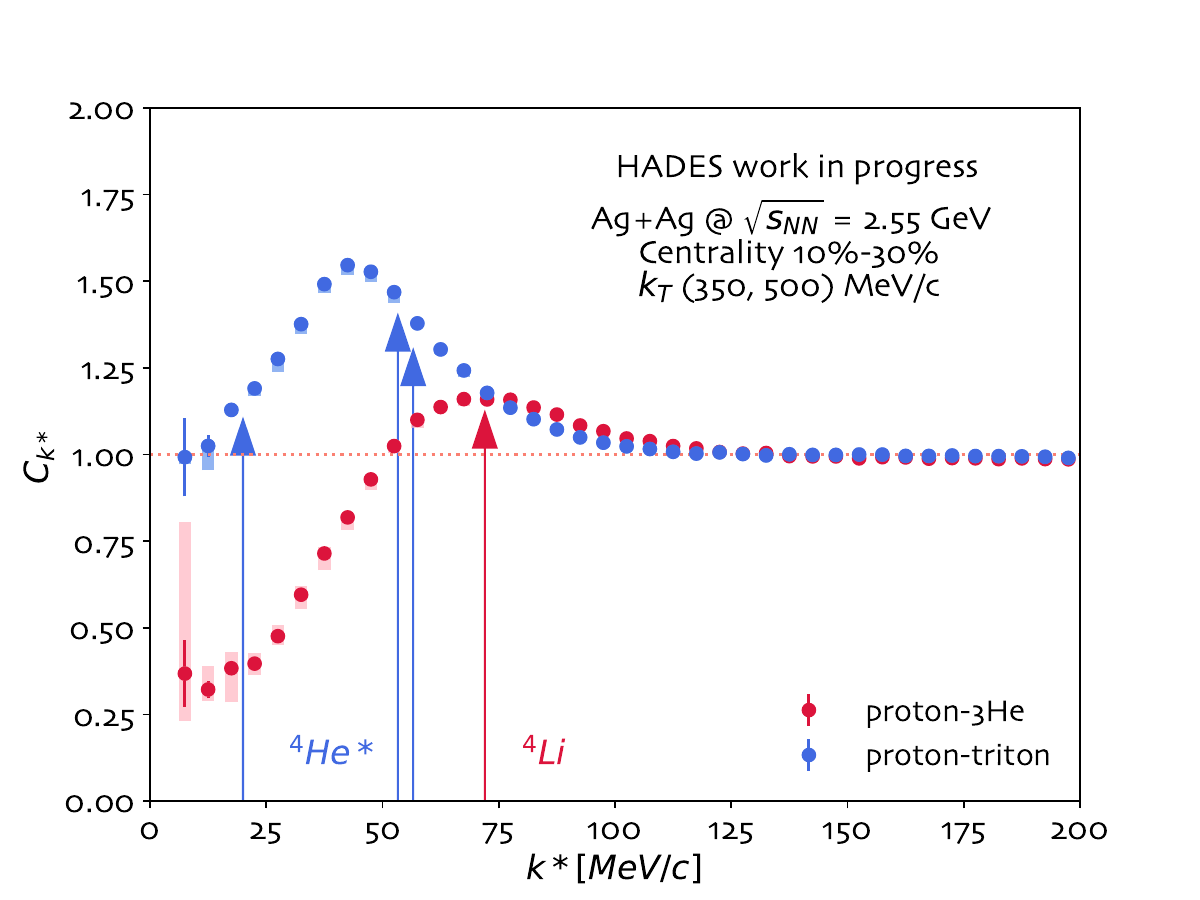}
\caption{Proton-$^3$He (red) and proton-triton (blue) correlation functions. Also depicted are the possible decays of $^4$He* and $^4$Li.}
\label{fig:decays}
\end{figure}

In Figure \ref{fig:decays} two CF are shown: p-t and p-$^3$He. Triton and $^3$He have similar masses and consist of three nucleons. However, their CF looks significantly different. The decays of excited states are causing the increase in CF values. The observed decays of these nuclear states are summarized below:\\
\textbf{$^4 He^{\star} \rightarrow p + t $:}
\begin{itemize}
    \item E = 20.21 MeV, $J_\pi$ = $0_+$, $\Gamma$ = 0.5 MeV, $\Gamma_p/\Gamma$ = 1, $k_1^{\star}$ = 20 MeV/c
    \item E = 21.01 MeV, $J_\pi$ = $0_+$, $\Gamma$ = 0.84 MeV, $\Gamma_p/\Gamma$ = 0.76, $k_2^{\star}$ =53.3 MeV/c
    \item E = 21.84 MeV, $J_\pi$ = 2, $\Gamma$ =2.01 MeV, $\Gamma_p/\Gamma$ = 0.63, $k_3^{\star}$ =  56.6 MeV/c
\end{itemize}
\textbf{$^4 Li \rightarrow p + ^{3}He$:}
\begin{itemize}
    \item $J_\pi$ = 2-, $\Gamma$ = 6.0 MeV,$\Gamma_p/\Gamma$ = 1, $k_1^{\star}$  72 MeV/c
\end{itemize}
Although the enhancements in the CF are clearly visible, the estimation of the number of decaying light nuclei is not trivial, as the shape of the CF is a convolution of the repulsive Coulomb interaction and atractive SI.

\section{Conclusions}
The study of proton-cluster CF, presented here, reveals the significant impact of the SI on the examined systems. The theoretical description proposed in CorAl is in good agreement with the data in the case of the p-p and p-d CF. However, the comparison of p-$^3$He CF shows discrepancies to the theoretical calculation in the region corresponding to the decay of $^4$Li. The extension of the CorAl calculations for p-d with p-wave and for p-$^3$He d-wave contributions is in progress. The investigation of $^3$He and triton correlations with protons shows that they are highly influenced by the decays of $^4$Li and $^4$He$^{\star}$. These studies give the unique opportunity to characterize the presence of these two nuclear states. More detailed investigations, including a full study of systematic uncertainties studies, is in progress.

\section{Acknowledgments}
This work was supported by Humboldt Foundation grant for postdoctoral fellows and U.S. Department of Energy grant DE-SC0020651.

\end{document}